\documentclass{article}
\usepackage{ijcai24}

\usepackage{times}
\usepackage{soul}
\usepackage{url}
\usepackage[hidelinks]{hyperref}
\usepackage[utf8]{inputenc}
\usepackage[small]{caption}
\usepackage{graphicx}
\usepackage{amsmath}
\usepackage{amsthm}
\usepackage{booktabs}
\usepackage{algorithm}
\usepackage{algorithmic}
\usepackage[switch]{lineno}

\usepackage{mathtools}
\usepackage{mathrsfs}
\usepackage{bm}
\usepackage{microtype}
\usepackage{makecell}
\usepackage{csquotes}
\usepackage{amsfonts}

\usepackage{wrapfig}
\usepackage{csquotes}
\usepackage{graphicx}
\usepackage{subcaption}
\usepackage{multirow}
\usepackage{xcolor}

\newtheorem{definition}{Definition}
\title{Reinforcement Nash Equilibrium Solver}

\author{
Xinrun Wang$^{1}$
\and
Chang Yang$^{2}$\and
Shuxin Li$^{1}$\and
Pengdeng Li$^{1}$\and\\
Xiao Huang$^2$\and
Hau Chan$^3$\And
Bo An$^1$\\
\affiliations
$^1$Nanyang Technological University, Singapore\\
$^2$The Hong Kong Polytechnic University, Hong Kong SAR, China\\
$^3$University of Nebraska-Lincoln, Lincoln, Nebraska, United States\\
\emails
\{xinrun.wang, shuxin.li, pengdeng.li, boan\}@ntu.edu.sg, chang.yang@connect.polyu.hk\\xiaohuang@comp.polyu.edu.hk, hchan3@unl.edu
}
\begin{document}

\maketitle

\begin{abstract}
Nash Equilibrium (NE) is the canonical solution concept of game theory, which provides an elegant tool to understand the rationalities. Though mixed strategy NE exists in any game with finite players and actions, computing NE in two- or multi-player general-sum games is PPAD-Complete. Various alternative solutions, e.g., Correlated Equilibrium (CE), and learning methods, e.g., fictitious play (FP), are proposed to approximate NE. For convenience, we call these methods as ``inexact solvers'', or ``solvers'' for short. However, the alternative solutions differ from NE and the learning methods generally fail to converge to NE. Therefore, in this work, we propose REinforcement Nash Equilibrium Solver (RENES), which \emph{trains a single policy to modify the games with different sizes and applies the solvers on the modified games where the obtained solution is evaluated on the original games}. Specifically, our contributions are threefold. i) We represent the games as $\alpha$-rank response graphs and leverage graph neural network (GNN) to handle the games with different sizes as inputs; ii) We use tensor decomposition, e.g., canonical polyadic (CP), to make the dimension of modifying actions fixed for games with different sizes; iii) We train the modifying strategy for games with the widely-used proximal policy optimization (PPO) and apply the solvers to solve the modified games, where the obtained solution is evaluated on original games. Extensive experiments on large-scale normal-form games show that our method can further improve the approximation of NE of different solvers, i.e., $\alpha$-rank, CE, FP and PRD, and can be generalized to unseen games. 
\end{abstract}

\section{Introduction}
Game theory provides a pervasive framework to model the interactions between multiple players~\cite{fudenberg1991game}. The canonical solution concept in non-cooperative games, i.e., the players try to maximize their own utility, is Nash Equilibrium (NE), where no player can change its strategy unilaterally to increase its own utility~\cite{nash1950equilibrium}. According to Roger Myerson, the introduction of NE is a watershed event for game theory and economics~\cite{myerson1999nash}. NE provides an impetus to understand the rationalities in much more general economic contexts and lies at the foundation of modern economic thoughts~\cite{myerson1999nash,goldberg2013complexity}. Mixed strategy NE exists in any game with finite players and actions~\cite{nash1950equilibrium}. However, from an algorithmic perspective, computing NE in two-player or multi-player general-sum games is PPAD-Complete~\cite{daskalakis2009complexity,chen2009settling}. In two-player zero-sum games, NE can be computed in polynomial time via linear programming. In more generalized cases, the Lemke–Howson algorithm is the most recognized combinatorial method~\cite{lemke1964equilibrium}, while using this algorithm to identify any of its potential solutions is PSPACE-complete~\cite{goldberg2013complexity}.

Given the difficulties of computing NE directly, there are many alternative solutions and learning methods proposed to approximate NE. For convenience, we call these solutions and learning methods as ``inexact solvers'', or ``solvers'' for short. Among them, one of the most widely used solution concepts is Correlated Equilibrium (CE)~\cite{aumann1974subjectivity,aumann1987correlated}, which considers the Bayesian rationality and is a more general solution concept than NE, i.e., any NE is CE. Recently, another solution concept $\alpha$-rank is proposed~\cite{omidshafiei2019alpha}, which adopts Markov-Conley Chains to highlight the presence of cycles in game dynamics and attempts to compute stationary distributions as a mean for strategy profile ranking. $\alpha$-rank is successfully applied to policy space response oracle (PSRO) to approximate NE~\cite{muller2020generalized}. The computations of both solution concepts are significantly simplified, but both solutions may be diverge from the accurate NE. 

To address the computational challenges, various heuristic methods are proposed to approximate NE. One of the most widely used methods is fictitious play~\cite{brown1951iterative,heinrich2015fictitious}, which takes the average of the best-responses iteratively computed by assuming the opponents play the empirical frequency policies. Another widely used learning method is projected replicator dynamics (PRD)~\cite{lanctot2017unified}, which is an advanced variant of replicator dynamics (RD)~\cite{taylor1978evolutionary} by enforcing explorations. Unfortunately, both methods fail to converge to NE in general cases. Another line of research is the homotopy method~\cite{herings2010homotopy,gemp2022sample}, which tries to build a continuum between the game and a simplified game with known NE and then uses the known NE to approximate the NE in the original game. The methods to build the continuum is important, which requires game-specific knowledge for efficient approximation. Furthermore, most previous methods require running the methods on each specific game and the obtained models are game-specific and lack the generalizability. Due to the limitation of space, we provide a detailed discuss of related works in Appendix B.

We leverage deep learning methods to improve the approximation of NE of the inexact solvers. One straightforward method is supervised learning (SL)~\cite{krizhevsky2012imagenet,he2016deep}, i.e., training a deep neural network which takes the game as input and outputs the NE directly. However, there are several issues of SL methods. First, it is difficult to generate the high quality dataset for SL methods due to: i) computing NE as labels is time-consuming and there are no efficient methods for computing NE, thus the generation of training datasets in SL methods is inefficient, ii) one game may have multiple NEs, i.e., multiple labels, and the computation of all equilibria and the equilibrium selection problem is not addressed~\cite{harsanyi1988general}. Second, the SL methods will inevitably generate the inexact solutions and further refinement methods are required to improve the approximation results, which makes the methods to be complicated. Therefore, inspired by~\cite{wang2021bi}, we leverage reinforcement learning (RL) methods, instead of using SL methods, to achieve this goal, i.e., \emph{iteratively modifying the games with different sizes with a single strategy learned by RL methods and applying the solvers, e.g., $\alpha$-rank, to the modified games to obtain the solutions which are evaluated on the original games}. However, there are two main issues: i) the games with different sizes result in varied inputs to the RL methods, where a simple neural network architecture, e.g., multi-layer perceptron (MLP) cannot handle, and ii) the number of payoff values grows exponentially with the number of players and actions, which brings difficulties to modify the games if we only change a payoff value once. 

To address the above issues, we propose REinforcement Nash Equilibrium Solver (RENES). Our main contributions are three-fold. First, we represent the games with different sizes as $\alpha$-rank response graphs, which are used to characterize the intrinsic properties of games~\cite{omidshafiei2020navigating}, and then leverage the graph neural network (GNN) to take the $\alpha$-rank response graphs as inputs. Second,  we use tensor decomposition, e.g., canonical polyadic (CP), to make the modifying actions fixed for games with different sizes, rather than changing a payoff value once. Third, we train the modifying strategy for games with the widely-used proximal policy optimization (PPO) and apply the solvers to solve the modified games, where the obtained solution is evaluated on original games. Extensive experiments on large-scale normal-form games, i.e., 3000 sampled games for training and 500 sampled games for testing, show that our method can further improve the approximation of NE of different solvers, i.e., $\alpha$-rank, CE, FP and PRD, and can be generalized to unseen games. To the best of our current knowledge, this work is the first effort in game theory that leverages RL methods to train a single strategy for modifying the games, and so as to improve the solvers' approximation performances.

\section{Preliminaries}
We present the preliminaries of game theory in this section. 

\paragraph{Normal-form Games.} Consider the $K$-player normal-form game, where each player $k\in[K]$ has a finite set of actions $\mathcal{A}^{k}$. We use $\mathcal{A}^{-k}$ to represent the action space excluding the player $k$, also for other terms. We denote the joint action space as $\mathcal{A}=\times_{k\in[K]}\mathcal{A}^{k}$. Let $\bm{a}\in\mathcal{A}$ be the joint action of $K$ players and $M(\bm{a})=\langle M^{k}(\bm{a})\rangle\in\mathbb{R}^{K}$ is the payoff vector of players when playing the action $\bm{a}$. A mixed strategy profile is defined as $\pi\in\Delta(\mathcal{A})$, which is a distribution over $\mathcal{A}$ and $\pi(\bm{a})$ is the probability that the joint action $\bm{a}$ will be played. The expected payoff of player $k\in[K]$ is denoted as $M^{k}(\pi)=\sum_{\bm{a}\in\mathcal{A}}\pi(\bm{a})M^{k}(\bm{a})$.

\paragraph{Solution Concepts.} Given a mixed strategy $\pi$, the best response of player $k\in[K]$ is defined as $\text{BR}^{k}(\pi)=\arg\max_{\mu\in\Delta(\mathcal{A}^{k})}[M^{k}(\mu, \pi^{-k})]$. A factorized mixed strategy $\pi(\bm{a})=\prod_{k\in[K]}\pi^{k}(a^{k})$ is Nash Equilibrium (NE) if $\pi^{k}\in\text{BR}^{k}(\pi)$ for $k\in[K]$. We define the NashConv value as $\texttt{NC($\pi$)}=\sum_{k\in[K]}M^{k}(\text{BR}^{k}(\pi), \pi^{-k})-M^{k}(\pi)$ to measure the distance of the mixed strategy from an NE. Computing NE in general-sum games is PPAD-Complete~\cite{daskalakis2009complexity}. Therefore, many alternative solutions are proposed to approximate NE. One of the most investigated alternative solutions is Correlated Equilibrium (CE)~\cite{aumann1987correlated}. A mixed strategy profile is CE if for all $k\in[K]$,
\begin{equation}
\sum\nolimits_{\bm{a}^{-k}\in\pi^{-k}} \pi(\bm{a})[M^{k}(\bm{a}) - M^{k}(\bm{a}^{-k}, b^{k})] \geq 0
\end{equation}
where $b^{k}\neq a^{k}$ and $b^{k}\in\mathcal{A}^{k}$. Any NE is a CE, therefore, CE is a generalized solution concept of NE, and CE is equivalent to NE in two-player zero-sum games. CE can be computed in polynomial time and many learning methods, e.g., regret matching~\cite{hart2000simple}, can lead to CE. As any NE is CE, CE also suffers the equilibrium problem, therefore, researchers propose many measures to select the equilibrium, e.g., maximum welfare CE and maximum entropy CE~\cite{marris2021multi}. $\alpha$-rank is recently proposed in~\cite{omidshafiei2019alpha}, which is the stationary distribution of the \emph{$\alpha$-rank response graph} constructed from the game. We will introduce the $\alpha$-rank response graph, as well as $\alpha$-rank in detail, in Section~\ref{sec:game_to_alpha_rank} as we also use the $\alpha$-rank response graph to transform the games into graphs. 

\paragraph{Learning Methods.} Instead of considering alternative solutions, many learning methods are proposed to approximate NE. Fictitious play (FP)~\cite{brown1951iterative,heinrich2015fictitious} is a famous method to approximate NE. FP starts with arbitrary strategies for players, and at each round, each player will compute the best response to the opponents' average behaviors. FP can converge to NE in certain classes of games, e.g., two-player zero-sum and many-player potential games~\cite{monderer1996fictitious}, while convergence is not guaranteed in general games. Replicator dynamics (RD)~\cite{schuster1983replicator} is a learning dynamics from evolutionary game theory, which is defined as  
\begin{equation}
    \dot{\pi}^{k}_{t}(a^{k})=\pi^{k}_{t}(a^{k})[M^{k}(a^{k}, \pi_{t}^{-k})-M^{k}(\pi_{t})],
\end{equation}
and RD can converge to NE under certain conditions. To ensure the exploration of RD, a variant of RD, projected replicator dynamics (PRD) is proposed in~\cite{lanctot2017unified}, which projects the strategy to the set $\Delta^{\gamma}(\mathcal{A}^{k})=\{\pi^{k}\in\Delta(\mathcal{A}^{k})|\pi^{k}(a^{k})\geq \frac{\gamma}{|\mathcal{A}^{k}|+1}, \forall a^{k}\in\mathcal{A}^{k}\}$. PRD is demonstrated to be more effective than RD to approximate NE in many games empirically~\cite{lanctot2017unified,li2021evolution}.

\section{Motivating Examples}
\label{sec:motivating_examples}
\begin{figure}[t]
\centering
\begin{subfigure}[b]{0.225\textwidth}
\centering
\includegraphics[width=0.975\textwidth]{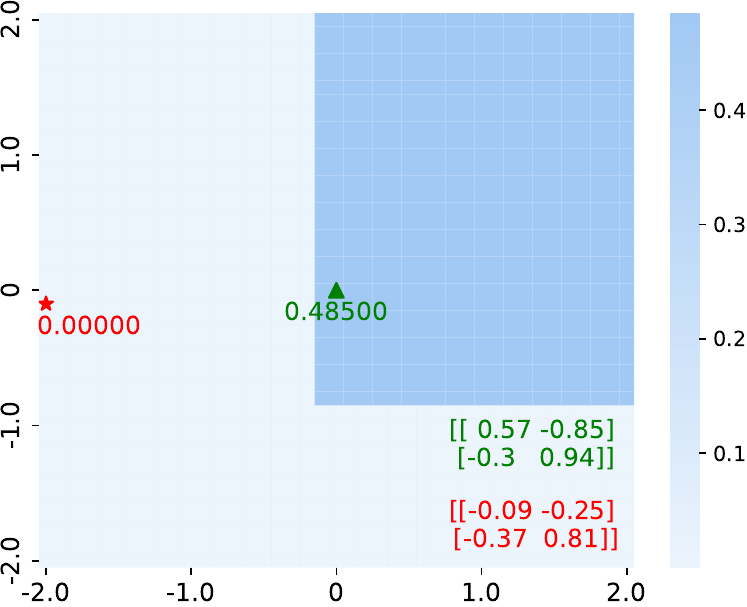}
\caption{$\alpha$-rank}
\label{fig:toy_alpharank}
\end{subfigure}
\begin{subfigure}[b]{0.225\textwidth}
\centering
\includegraphics[width=0.975\textwidth]{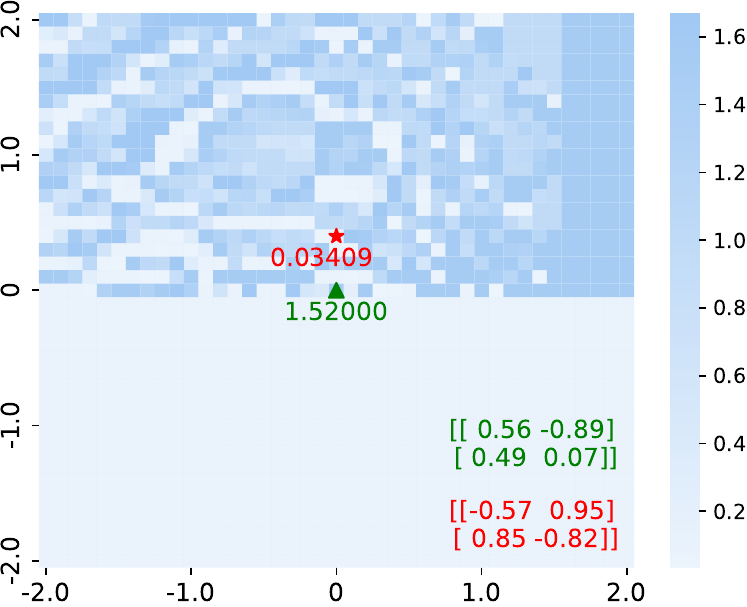}

\caption{CE}
\label{fig:toyl_ce}
\end{subfigure}
\begin{subfigure}[b]{0.225\textwidth}
\centering
\includegraphics[width=0.975\textwidth]{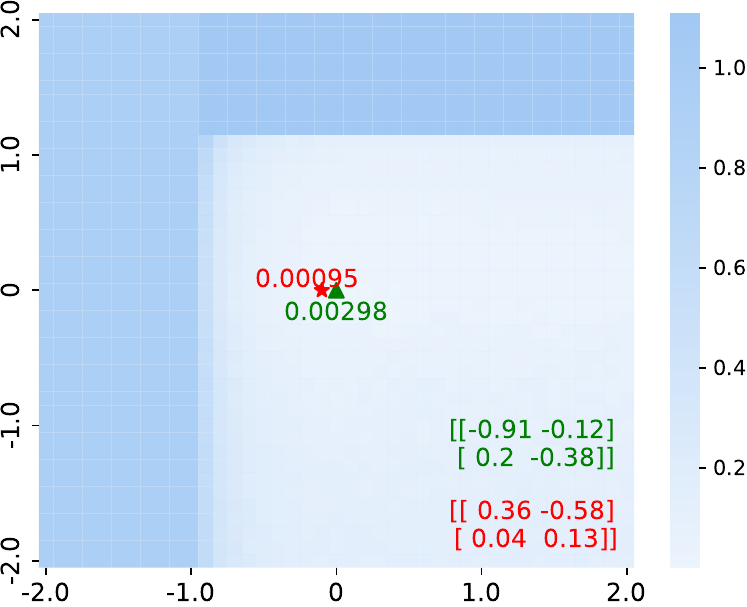}
\caption{FP}
\label{fig:toy_fp}
\end{subfigure}
\begin{subfigure}[b]{0.225\textwidth}
\centering
\includegraphics[width=0.975\textwidth]{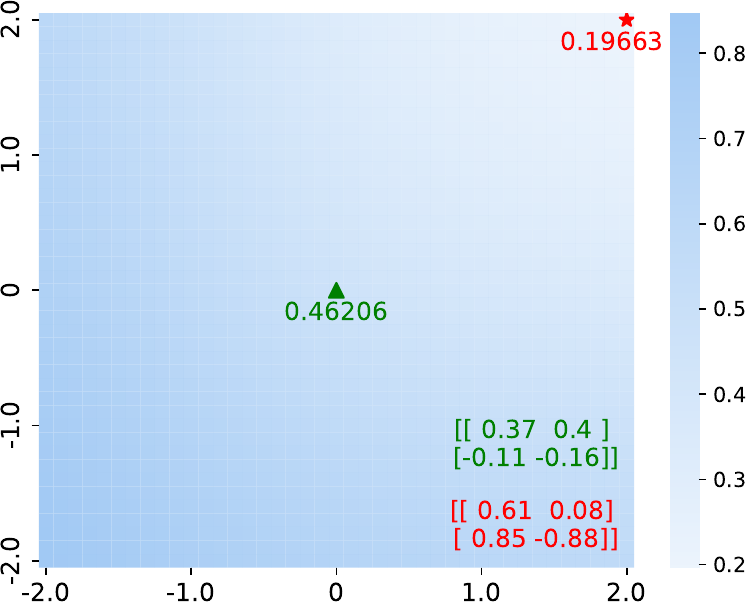}

\caption{PRD}
\label{fig:toy_prd}
\end{subfigure}
\caption{Motivating Examples. The payoff matrices of the two players are displayed in the figure in green and red colors, respectively. The x-axis and y-axis are the values of $\delta^{1}$ and $\delta^{2}$, respectively. For plotting, the interval $[-2, 2]$ is discredited with a step size 0.1. The NashConv values of the solvers on the original games $M$ is marked as the green triangle and the minimal NashConv value is marked as the red star. We note that the games $M$ selected for the solvers are specifically designed and differ from each other.}
\label{fig:toy}
\end{figure}
In this section, we provide motivating examples of RENES to demonstrate that the strategies obtained by alternative solutions and learning methods can diverge from NE strategies and can be improve through modifying the games.

Specifically, we consider the general -sum games with 2 players and each player has 2 actions. Thus, the game can be represented as a $2\times2\times2$ tensor $M=\langle M^{1}, M^{2}\rangle$ where $M^{k}=\left[\begin{smallmatrix} m_{11}^{k} & m_{12}^{k} \\ m_{21}^{k} & m_{22}^{k} \end{smallmatrix}\right]$, $k\in\{1, 2\}$ and the payoff values are in $[-1, 1]$. For simplicity and easy visualization, we only modify the payoff values of the first actions of players. Denoting the modification values as $\delta^{1}$ and $\delta^{2}$, therefore, the modified game is  $\tilde{M}=\langle \tilde{M}^{1}, \tilde{M}^{2}\rangle$ where $\tilde{M}^{k}=\left[\begin{smallmatrix} m_{11}^{k}+\delta^{k} & m_{12}^{k} \\ m_{21}^{k} & m_{22}^{k} \end{smallmatrix}\right]$, $k\in\{1, 2\}$. The values of $\delta^{k}\in[-2, 2]$. We then apply the considered solvers, i.e., alternative solutions and learning methods, to the modified games and evaluate on the original games. 

The results are displayed in Figure~\ref{fig:toy}. We can observe that in the specifically designed games, the solvers diverge from the exact NE. Through only modifying the payoff values of the first actions, we can significantly improve the approximation of $\alpha$-rank, CE and PRD and also slightly improve on FP. With larger spaces of the modification, we can even further improve the approximation of solvers. The results in Figure~\ref{fig:toy} demonstrate the effectiveness of the fundamental idea of RENES, i.e., improving the approximation of solvers through modifying the games. However, there are still several issues: i) when the space of modification is larger, the enumeration of all possible combinations is impossible, ii) for different games with different sizes, the modification spaces vary significantly. Therefore, we propose RENES which can provide the efficient modification strategies for the games with different sizes.

\section{RENES}

\begin{figure}[ht]
% \vspace{-5pt}
\centering
\includegraphics[width=0.28\textwidth]{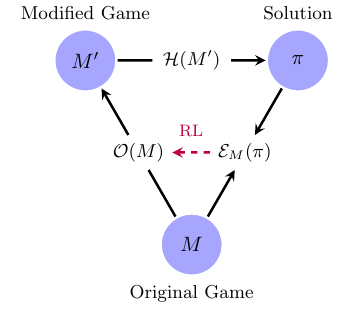}
\caption{Flow of RENES. Specifically, staring with the original game $M$, the modification oracle $\mathcal{O}$ modified the game to $M'$ and the solver $\mathcal{H}$ is applied to the modified game $M'$. The obtained solution $\pi$ is evaluated on $M$ with $\mathcal{E}$.}
\label{fig:flow_renes}
\end{figure}

In this section, we introduce the proposed REinforcement Nash Equilibrium Solver (RENES). The general procedure of RENES is displayed in Figure~\ref{fig:flow_renes}. Specifically, RENES is formed with three components: i) a modification oracle $\mathcal{O}$, represented as neural networks and trained with RL methods, e.g., PPO~\cite{schulman2017proximal}, which takes the original game $M$ as input and generates the modified game $M'$, ii) an inexact solver $\mathcal{H}$, which takes the modified game $M'$ and generates the solution $\pi$, and iii) an evaluation measure $\mathcal{E}$, which evaluates the obtained solution $\pi$ on the original game $M$ and provides the reward signal to the RL method when training the modification oracle $\mathcal{O}$. The solvers considered in this paper are $\alpha$-rank, CE, FP, PRD and the evaluation measure used in this paper is NashConv~\cite{muller2020generalized}. We will provide the details of the design and training of the modification oracle $\mathcal{O}$ in the rest of this section.

\subsection{Games as $\alpha$-Rank Response Graphs}
\label{sec:game_to_alpha_rank}

The normal-form games are normally represented as high-dimensional tensors. However, current deep learning methods cannot efficiently handle high dimensional tensors. There are two possible solutions: i) flattening the tensor into a 1-D vector and using MLP as the modification oracle $\mathcal{O}$, where the relations between payoff values will be eliminated and the length of the vector is varied for games with different sizes, and ii) using convolutional neural network (CNN), which is efficient to process images~\cite{he2016deep}. However, CNN still cannot process high dimensional tensors, e.g., 4-D, and cannot handle games with different sizes. Another option is that we can add a cap of the game size and use the maximum size of the game to build the policy network (we use 0 to fill in the tensor if the game is smaller than the maximum size during training and testing), however, this will still hurt the generalizability of RENES, as it cannot handle the games beyond the maximum size.

\begin{figure}[ht]
\centering
\begin{subfigure}[b]{0.2\textwidth}
\centering
\begin{small}
\begin{tabular}{c|c|c|c|}
    \multicolumn{1}{c}{} & \multicolumn{1}{c}{R} & \multicolumn{1}{c}{P} & \multicolumn{1}{c}{S} \\
    \cline{2-4}
    R & $0$  & $-1$& $1$ \\
    \cline{2-4}
    P & $1$& $0$& $-1$ \\
    \cline{2-4}
    S & $-1$& $1$ & $0$\\
    \cline{2-4}
\end{tabular}
\end{small}
\caption{Payoff table}
\label{fig:rps_payoff}
\end{subfigure}
\quad
\begin{subfigure}[b]{0.25\textwidth}
\centering
\includegraphics[width=0.87\textwidth]{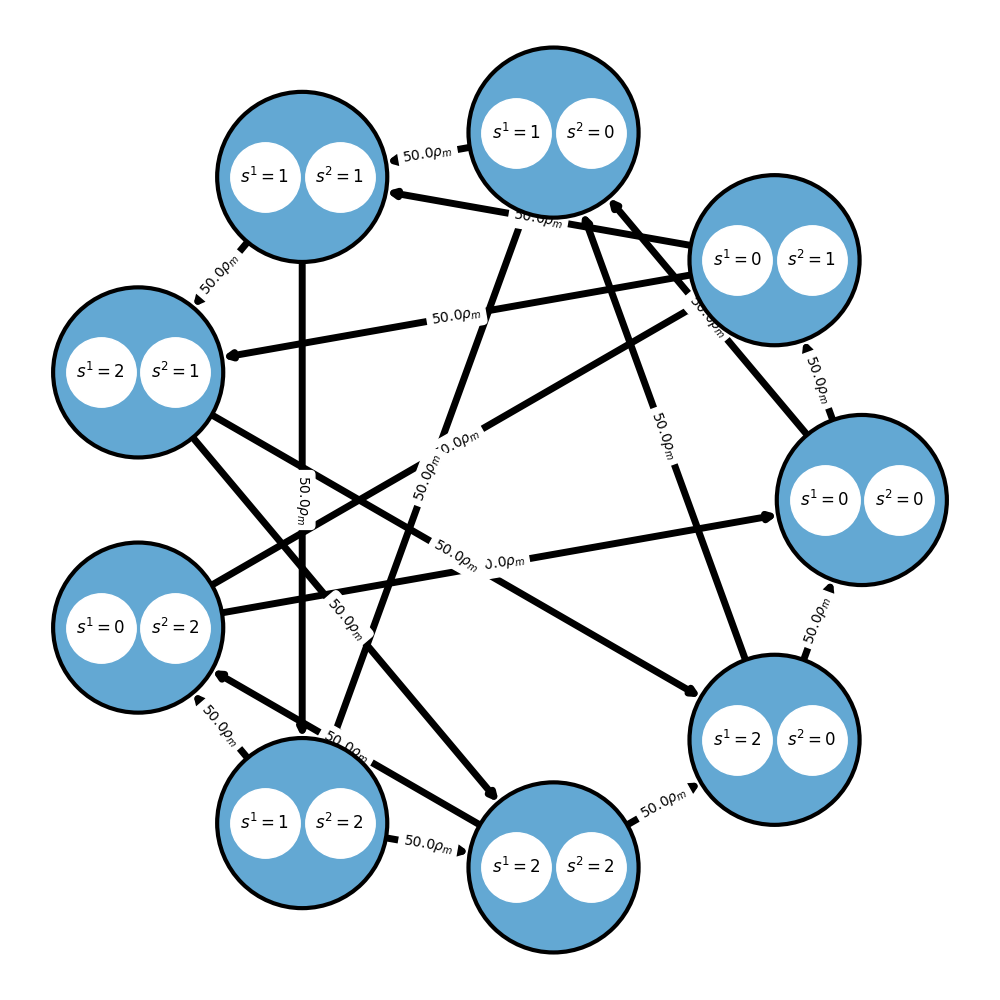}
\label{fig:rps_alpharank}
\caption{$\alpha$-rank response graph}
\end{subfigure}
\caption{Payoff table and $\alpha$-rank response graph for Rock-Paper-Scissors (RPS) game, where $\alpha=100$ and $m=50$ when computing the $\alpha$-rank response graph.}
\label{fig:rps}
\end{figure}

To handle the games with different sizes, we represent the games as the $\alpha$-rank response graphs, which is shown to represent the intrinsic properties of games in~\cite{omidshafiei2020navigating}, and then use graph neural network (GNN)~\cite{kipf2016semi,velivckovic2018graph,yun2019graph} to extract the features of games. We note that GNN can efficiently handle the graphs with different sizes~\cite{li2018combinatorial}, as it takes the neighboring information to update the node embeddings. 
The definition of $\alpha$-rank response graph, as well as $\alpha$-rank, is defined in Definition~\ref{def:alpha_rank_respnse_graph}, where each joint action corresponds to a node on the graph, and the two nodes are connected if there is only one player has different action in the joint actions. The transition probabilities are determined by Eq.~(\ref{eq:alpha_rank_prob}). A concrete example of $\alpha$-rank response graph for the rock-paper-scissors game is displayed in Figure~\ref{fig:rps}. After representing the games as graphs, we use the same value, e.g., 1.0, as the node features and the transition probabilities as the weights of edges. Then, we use GNN, e.g., GCN~\cite{kipf2016semi} to process the $\alpha$-rank response graph to obtain the embedding of the graph, and then we add an MLP to generate the outputs. We use PPO to train the modification oracle, where both actor and critic use this architecture to generate the action and the state-action value, respectively. More training details can be found in Section~\ref{sec:training_renes}. 

\begin{definition}
\label{def:alpha_rank_respnse_graph}
The $\alpha$-rank response graph is defined over all joint pure actions, specified by a transition matrix $C$. Given a joint pure action $\bm{a}$ and $\bm{a}'=\langle \sigma^{k}, \bm{a}^{-k}\rangle$ is a joint pure action which equals to $\bm{a}$ except the player $k$, we denote the transition probability from $\bm{a}$ to $\bm{a}'$ as $C(\bm{a}, \bm{a}')$ 
% \begin{small}
\begin{equation}
\label{eq:alpha_rank_prob}
\begin{cases}
\eta\frac{1-\exp(-\alpha (M^{k}(\bm{a}')-M^{k}(\bm{a})))}{1-\exp(-\alpha m (M^{k}(\bm{a}')-M^{k}(\bm{a})))}, & \text{if } M^{k}(\bm{a})\neq M^{k}(\bm{a}') \\
\frac{\eta}{m}, & \text{otherwise}
\end{cases}
\end{equation}  
% \end{small}
where $\eta=1/(\sum_{k\in[K]}(|\mathcal{A}^{k}|-1))$. The self-transition probability is defined as $C(\bm{a}, \bm{a})=1-\sum_{k\in[k], \bm{a}'|\sigma^{k}\in\mathcal{A}^{k}} C(\bm{a}, \bm{a}')$. If two joint actions $\bm{a}$ and $\bm{a}'$ differ in more than one player's action, $C(\bm{a}, \bm{a}')=0$. The values of $\alpha$ and $m$ are the selection pressure and the number of populations, respectively, which are specified by users. The transition matrix $C$ define a directed graph, i.e., $\alpha$-rank response graph, where the stationary distribution of $C$ is the $\alpha$-rank distribution. 
\end{definition}

\subsection{Action Spaces of RENES}

Most RL methods are designed for the problems with fixed action spaces. However, when the game sizes change, the number of elements which can be changed is also different. Therefore, the action space of RENES needs to be specially designed. 
A naive definition of the action space is that at each step, RENES changes the payoff of a player in a specific joint action, i.e., generating the indices of the elements in the payoff table and the way to modify this payoff value. In this case, the action size of RENES is $K\cdot\prod_{k\in[K]}|\mathcal{A}^{k}|$, which grows exponentially along with the number of actions of each player. On the other hand, we still need to specify the maximum size of the game sizes to generate valid indices of the elements, which hurts the generalizability of RENES. 

Therefore, we consider a more compact action space with tensor decomposition~\cite{kolda2009tensor}. Specifically, we use the canonical polyadic (CP) decomposition of the payoff table $M$ and set the rank $r$ to be fixed and the action of RENES is the coefficients over $r$:
\begin{equation}
    M\approx\sum\nolimits_{i=1}^{r}\lambda_{i} \cdot m_{1, i}\otimes m_{2, i}\otimes\cdots\otimes m_{K+1, i},
\end{equation}
where $\bm{\lambda}=\langle\lambda_{i}\rangle, i=1,\cdots, r$ are the weights of the decomposed tensors and $m_{k,i}, k\in\{1, \dots, K+1\}$ are the factors which are used to modify the game. For the decomposition, the weight $\bm{\lambda}=\bm{1}$\footnote{The tensor decomposition is implemented by \texttt{TensorLy} (\url{https://github.com/tensorly/tensorly}). Other implemented decomposition methods can also be used.}. Given any arbitrary weight $\bm{\lambda}$, we can reconstruct the payoff tensor with the reconstruction oracle $\mathcal{R}_{M}(\bm{\lambda})$. Therefore, we let the modified oracle $\mathcal{O}$ to modify the weights and update the game by 
\begin{equation}
\label{eq:game_modification}
    M_{t}=M_{t-1} + \eta\cdot \mathcal{R}_{M}(\bm{\lambda}).
\end{equation}
With the tensor decomposition, we can use a fixed size of action space of RENES, specified by $r$. The tensor decomposition can be viewed as a simple method of the abstraction~\cite{brown2015simultaneous,brown2017safe}, and more sophisticated and decomposition methods can be considered in future works~\cite{burch2014solving}.
% We will apply the heuristic methods or compute the alternative solutions of $M_{t}$. In this case, we can make the action size of the modified oracle to be independent of the game size.
\subsection{MDP Formulation of RENES}

As RL relies on the Markov Decision Process (MDP) formulation~\cite{schulman2017proximal}, we will also first reformulate our problem into an MDP:
\begin{itemize}
    \item \textbf{States.} The state of RENES is defined as the tuple with the original game $M_{0}$ and the current game $M_{t-1}$. Note that we compute the solution with the current game $M_{t-1}$ and evaluate the obtained solution on the original game $M_{0}$, therefore, $\langle M_{0}, M_{t-1}\rangle$ give the full information of the underlying MDP to solve, therefore, we ignore all the intermediate games generated, i.e., $M_{t}, t\in\{1, \dots, t-2\}$, which can simplify and stabilize the training.
    \item \textbf{Actions.} The actions of RENES are the weights $\bm{\lambda}$ over the decomposition factors, which will be used to update $M_{t-1}$ to obtain $M_{t}$ following Eq. (\ref{eq:game_modification}). 
    \item \textbf{Transition Function.} After obtaining the new modified game $M_{t}$, the problem will transit to the new state defined by the tuple of $M_{0}$ and $M_{t}$.
    \item \textbf{Reward Function.} The immediate reward at step $t$ is defined as $\texttt{NC}(\pi_{t-1})-\texttt{NC}(\pi_{t})$, where $\pi_{t}$ is the strategy obtained when applying the solver to the modified game $M_{t}$. As we use NashConv (lower is better) as our evaluation measure, the decrement of the NashConv will be the positive reward of RENES. With maximizing the accumulated reward, RENES can boost the approximation of NE of the solver. We also consider using a normalized NashConv measure, defined as $[\texttt{NC}(\pi_{t-1})-\texttt{NC}(\pi_{t})]/\texttt{NC}(\pi_{0})$, where $\texttt{NC}(\pi_{0})$ is the normalizer. This normalized NashConv can make the performances on different games comparable. However, when $\texttt{NC}(\pi_{0})$ is small, e.g., less than $0.01$, and  $|\texttt{NC}(\pi_{t-1})-\texttt{NC}(\pi_{t})|$ is much larger than $\texttt{NC}(\pi_{0})$, e.g., $1.0$, this can make the reward value be extremely large and cannot be used for training. Therefore, the normalized NashConv is only used for evaluation.
    \item \textbf{Horizon \& Discounting Factor.} The horizon $T$ specifies the maximum step, e.g., 50, where the modification can be used to modify the game for better performance. The discounting factor is denoted as $\gamma\in (0, 1]$. 
\end{itemize}

\subsection{Training RENES with PPO}
\label{sec:training_renes}
After reformulating the training of RENES as MDP, we then train RENES with RL methods. RL is an area of the policy optimization in complex sequential decision-making environments~\cite{sutton2018reinforcement}. RL methods rely on the trail-and-error process to explore the solution space for better policies. The primary RL method is Q-learning~\cite{watkins1992q,mnih2015human}, which can only be used on the problems with discrete actions, and the policy gradient methods are proposed for the problems with both discrete and continuous actions~\cite{sutton1999policy,mnih2016asynchronous,haarnoja2018soft}.

PPO is an on-policy policy gradient method, which is a simplified, but more data efficient and reliable, variant of Trust Region Policy Optimization (TRPO)~\cite{schulman2015trust}, which leverages the ``trust region'' to bound the update of the policy to avoid training collapse. Compared with TRPO, PPO is more data efficient and with more reliable performances than TRPO, while only using the first-order optimization for computational efficiency.

Specifically, PPO is maximizing the objective 
\begin{equation}
  J(\theta)  =\mathbb{E}\left[\min(r_{\theta}\cdot A, \texttt{clip}(r_{\theta}, 1-\epsilon, 1+\epsilon)\cdot A)\right],
\end{equation}
where $r_{\theta}$ is the importance sampling ratio conditional on $\theta$, $\theta$ is the parameter of the policy, $A$ is the advantage value which is computed by using the discounted accumulative reward minus the critic network prediction of the state-action value, and $\epsilon$ is the hyperparameter which controls the boundary of the trust region. We only provide a short introduction of PPO in this section, as we take PPO as a blackbox for optimizing the modification oracle $\mathcal{O}$. Other RL methods, e.g., soft actor critic (SAC)~\cite{haarnoja2018soft}, can be used and for more details of RL, we refer readers to~\cite{sutton2018reinforcement}.

\section{Experiments}
In this section, we present the experimental results of RENES on large-scale normal-form games. We consider two cases: i) \textbf{simple case} where all games have the same size to verify the idea of modifying the games to boost the performance of inexact solvers, and ii) \textbf{general case} where the games have different sizes to verify that the design of RENES can handle the game with different sizes. We then conduct the ablation study on the numbers of the action dimensions and the horizon.

\subsection{Experimental Setups}
% % We present the experimental setups in this section.
\paragraph{Games and Solvers.} For the games, we randomly sample 3000 games for training, and 500 games for testing to verify the ability of Renes to generalize to unseen games. The games are represented as high dimensional payoff tensors with dimensions as $[K, |\mathcal{A}^{1}|, \dots, |\mathcal{A}^{K}|]$, which are used for tensor decomposition. We also do the normalization of the payoff to stablize the training. The four selected solvers are: $\alpha$-rank, CE, FP and PRD. For $\alpha$-rank and PRD, we use the implementations in OpenSpiel~\cite{lanctot2019openspiel}\footnote{\url{https://github.com/deepmind/open_spiel}}. We use regret matching~\cite{hart2000simple} in OpenSpiel for CE and implement FP by ourselves. More details of the games and solvers are in Appendix~B.

\paragraph{Evaluation Measures.} Different from RL where the accumulated reward is considered, we focus on the minimum NashConv during the $T$ steps of the modification. As the solvers have different performances on different games, we would use the normalized NashConv measure to make the performances comparable. Specifically, the performance of RENES on a specific game is measured by $1-\frac{\min_{t\in[T]}\{\texttt{NC}(\pi_{t})\}}{\texttt{NC}(\pi_{0})}$, which measures the relative improvement over the performance of the solver on the original game, therefore, $0$ implies no improvement and $1$ implies the obtained solution is the exact NE. We take the average over all games to measure the performance. 

\paragraph{Baselines.} We take the two baselines to compare with RENES: i) the performance of the solver on the original games without any modification, ii) the performance of the random policy of modifying games. The two baselines correspond to the value of 0 and the performance at 0 step of the RL training using the proposed measure. Most of the previous methods require running the methods on specific games and the learned policies cannot be generalized to other games, while RENES can generalize to unseen games without training, so we do not consider these SoTA methods as our baselines.\footnote{More justifications of the baselines can be found in Appendix A.}

\paragraph{Training with PPO} For the training, we set the decomposition rank $r=10$, i.e., the number of the action dimensions is 10, and $T=50$, i.e., the number of the maximum steps of the modification is 50. Conceptually, the more action dimensions and the more steps, the better performance of RENES. Larger values of $r$ may bring the difficulties of training and make the training unstable. As at each time step, we need to run solvers to solve the modified game, which is time-consuming when $T$ is very large. We choose these two values to balance the trade-off between the performance and the efficiency. We run the experiments with three seeds. The values of the hyperparameters can be found in Appendix B. Due to the limitation of the computational resources, we do not conduct the exhaustive tuning of the hyperparameters.

\subsection{Simple Case}

\begin{figure}
\centering
\begin{subfigure}[b]{0.225\textwidth}
\centering
\includegraphics[width=0.975\textwidth]{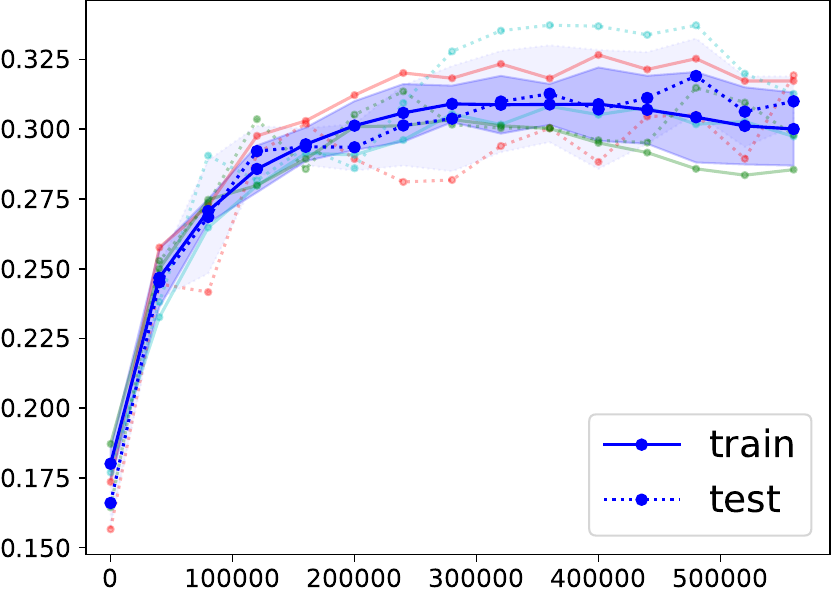}
\caption{$\alpha$-rank}
\label{fig:simple_alpharank}
\end{subfigure}
\begin{subfigure}[b]{0.225\textwidth}
\centering
\includegraphics[width=0.975\textwidth]{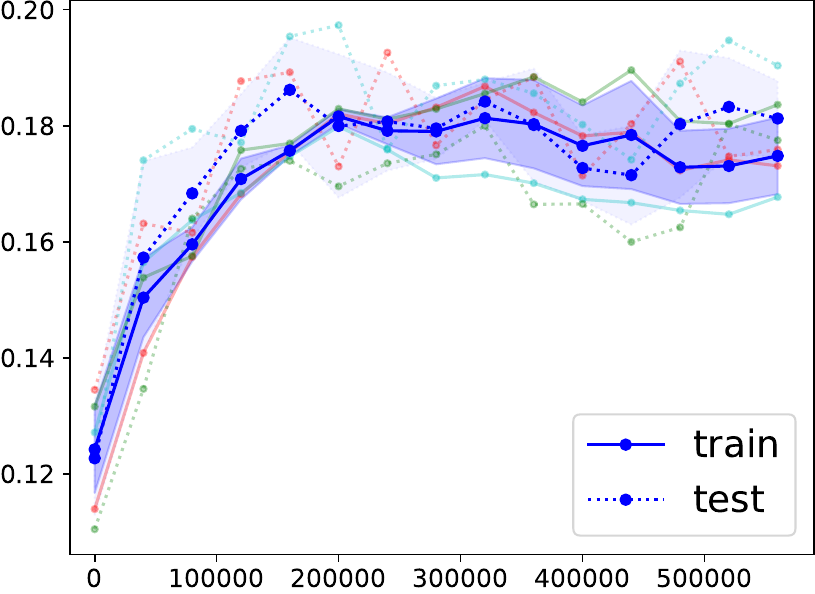}
\label{fig:simple_ce}
\caption{CE}
\end{subfigure}
\begin{subfigure}[b]{0.225\textwidth}
\centering
\includegraphics[width=0.975\textwidth]{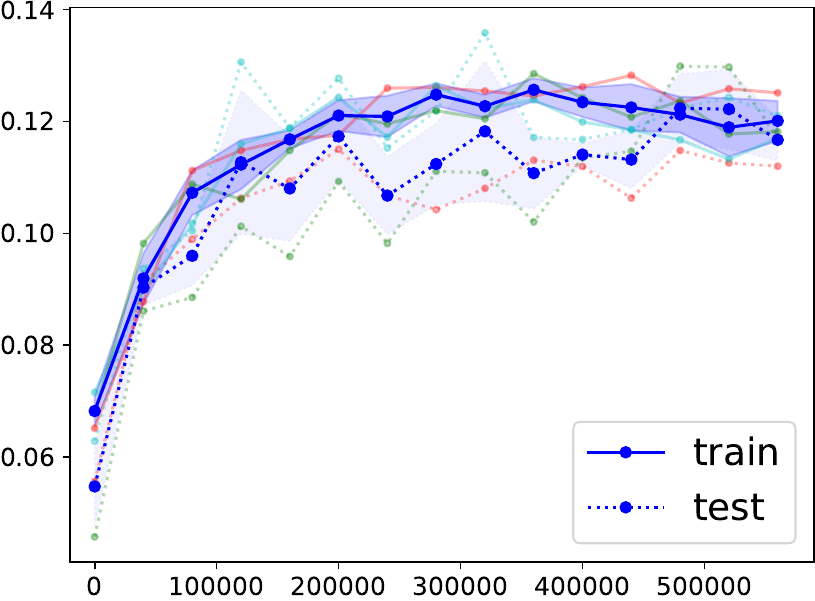}
\caption{FP}
\label{fig:simple_fp}
\end{subfigure}
\begin{subfigure}[b]{0.225\textwidth}
\centering
\includegraphics[width=0.975\textwidth]{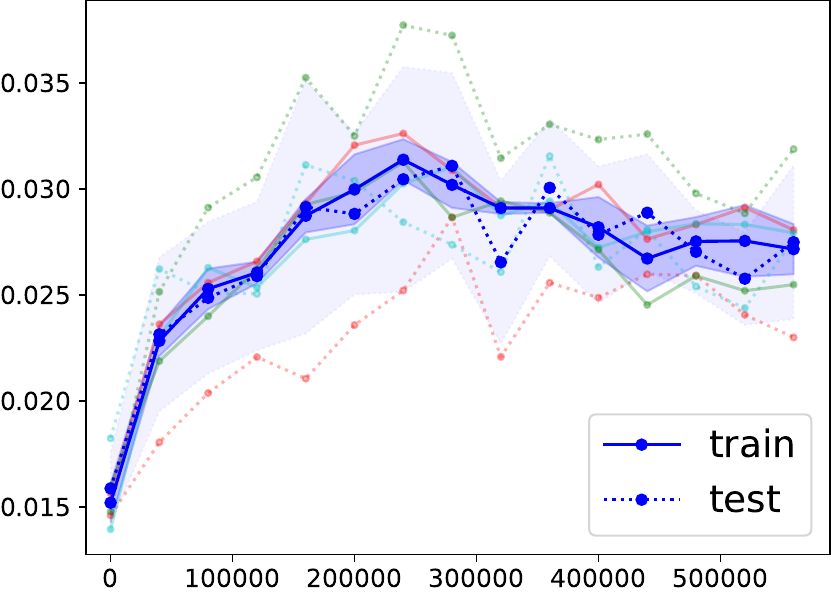}
\caption{PRD}
\label{fig:simple_prd}
\end{subfigure}
\caption{Results of RENES in simple case. The solid lines and dotted lines are the results on the training set and the testing set of games, respectively. The transparent lines are the results with different seeds, as the runs will different seeds achieve the best performances in different epochs, so we plot them for better understanding of the training across different seeds. The blue line is the averaged results and the shaded area plots the standard deviation. Note that the y-axis scale differs across figures for better visualizations. The same style is also adopted in Figures~\ref{fig:general} and \ref{fig:simple_ablation}.}
\label{fig:simple}
\end{figure}
In this section, we present the experiments on the simple case. For the simple case, we randomly sample 3000 games for training and 500 games for testing where all games are 2-player games with 5 actions of each player. As all games have the same size, we simply flatten the payoff tensors to 1-D vectors and using MLP for the policy and critic in PPO.

\begin{table*}[ht]
\centering
\caption{Results of RENES in Simple Case. To calculate the values, we pick the best values across different epochs for each seed and compute the mean values and standard deviation values. }
\label{tab:simple}
\begin{tabular}{c|c|cccc}
\toprule
     &&  $\alpha$-rank & CE & FP & PRD\\\midrule
\multirow{2}*{Training}&Random  & 0.180(0.006)& 0.124(0.007) & 0.068(0.003)&0.015(0.001)\\
&RENES & 0.313(0.010)& 0.185(0.004) & 0.128(0.001)& 0.032(0.001)\\
\midrule
\multirow{2}*{Testing}&Random  & 0.180(0.006)& 0.123(0.010) & 0.055(0.007)&0.016(0.002)\\
&RENES & 0.324(0.010)& 0.190(0.007) & 0.127(0.009)& 0.033(0.004)\\
\bottomrule
\end{tabular}  

\end{table*}

The results of the simple case are displayed in Figure~\ref{fig:simple} and Table~\ref{tab:simple}. We observe that the training on the training set of games can be generalized to the testing set, which indicates that RENES can be used as a general policy to modify the games, even on unseen games. More specifically, we observe that RENES can significantly boost the performance of $\alpha$-rank, i.e., larger than $0.3$ over all three seeds, as shown in Figure~\ref{fig:simple_alpharank}, and achieve $0.313$ and $0.324$ on training and testing sets, respectively. For PRD, both random policy and RENES can only bring small improvements, i.e., smaller than $0.05$ over all seeds, as shown in Figure~\ref{fig:simple_prd}, and only achieve $0.032$ and $0.033$ on training and testing sets, respectively. For the other solvers, RENES can bring notable improvements, i.e., larger than $0.16$ for CE and $0.1$ for FP over three seeds. Overall, we can conclude that the performances of inexact solvers can be boosted through modifying the games. 
We also observe that longer training of RENES does not necessarily improve the performance, i.e., Figures~\ref{fig:simple_alpharank}, \ref{fig:simple_fp} and \ref{fig:simple_prd}, as observed in other RL experiments. We believe that with more tuning of the hyperparameters, RENES can achieve better performances.

\subsection{General Case}

We then conduct the experiments on the general case. In the general case, we also sample 3000 games for training and 500 games for testing, where the games have $\{2, 3\}$ players and each player is with $\{2, 3, 4\}$ actions. We only focus on small games as the running of solvers for large multi-player is even more time-consuming. As the game sizes vary in this case, we will take the GNN as the base network and add an MLP head to form the policy and the critic in PPO, respectively. Other settings are the same as the simple case.

\begin{figure}[ht]
\centering
\begin{subfigure}[b]{0.225\textwidth}
\centering
\includegraphics[width=0.975\textwidth]{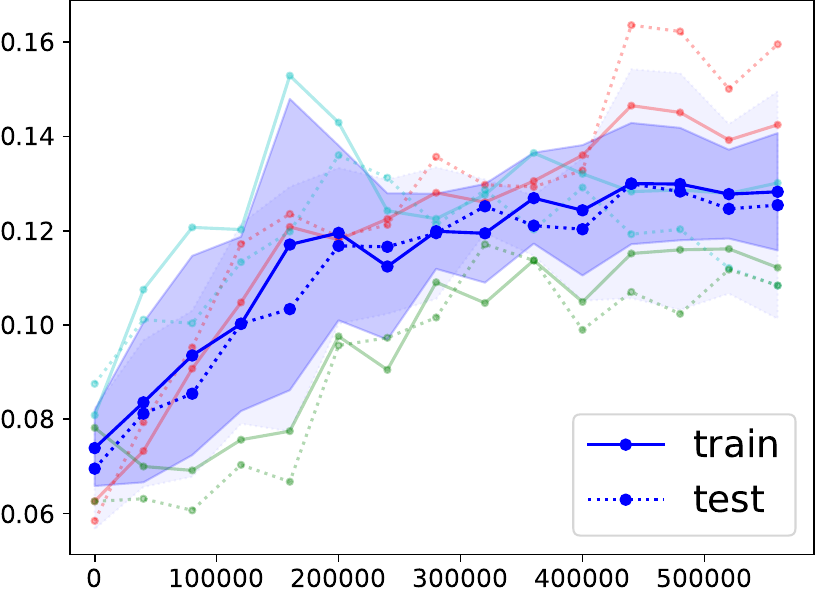}
\caption{$\alpha$-rank}
\label{fig:general_alpharank}
\end{subfigure}
\begin{subfigure}[b]{0.225\textwidth}
\centering
\includegraphics[width=0.975\textwidth]{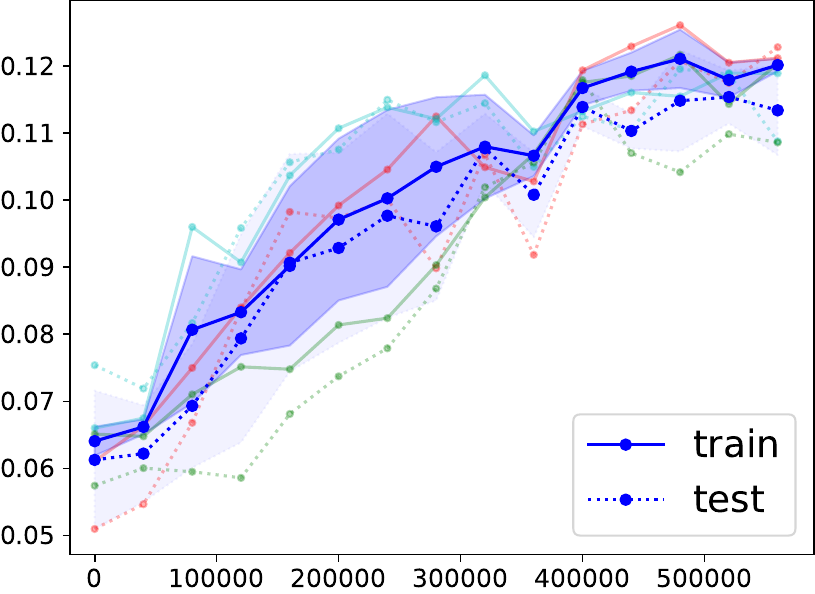}

\caption{CE}
\label{fig:general_ce}
\end{subfigure}
\begin{subfigure}[b]{0.225\textwidth}
\centering
\includegraphics[width=0.975\textwidth]{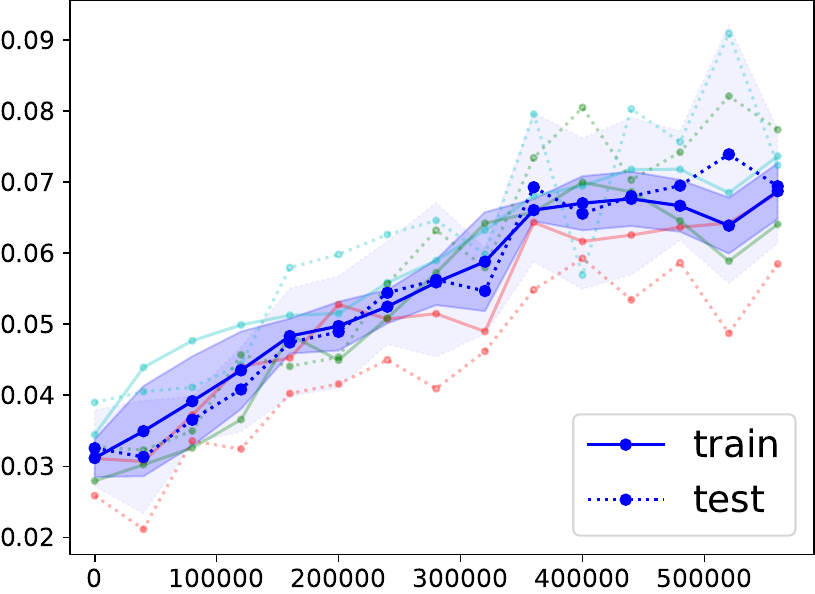}
\caption{FP}
\label{fig:general_fp}
\end{subfigure}
\begin{subfigure}[b]{0.225\textwidth}
\centering
\includegraphics[width=0.975\textwidth]{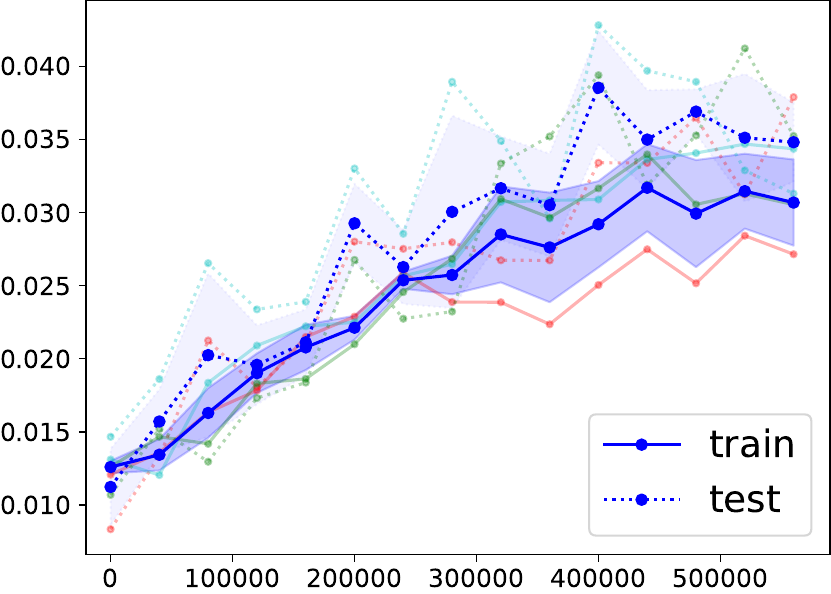}

\caption{PRD}
\label{fig:general_prd}
\end{subfigure}
\caption{Results of RENES in general case}
\label{fig:general}
\end{figure}

The results of the general case are displayed in Figure~\ref{fig:general} and Table~\ref{tab:general}. We observe that similar to the simple case, the policy of RENES trained on the training set can be generalized to the testing set, which indicates that RENES can be a general policy for unseen games even when the game sizes vary. Compared with the simple case, RENES achieves lower performances over different solvers on the general case. Specifically, RENES still brings the largest improvement for $\alpha$-rank, i.e., $0.139$ on both training and testing dataset, and the smallest improvement for PRD, i.e., $0.032$ and $0.041$ on training and test datasets, respectively. For the other two solvers, RENES also brings notable improvements, i.e., larger than $0.120$ and $0.071$ on both training and test datasets, respectively.

\begin{table*}[ht]
\centering
\caption{Results of RENES in General Case. }
\label{tab:general}
\begin{tabular}{c|c|cccc}
\toprule
     &&  $\alpha$-rank & CE & FP & PRD\\\midrule
\multirow{2}*{Training}&Random  & 0.074(0.008)& 0.064(0.002) & 0.031(0.003)&0.013(0.001)\\
&RENES & 0.139(0.016)& 0.122(0.003) & 0.071(0.002)& 0.032(0.003)\\
\midrule
\multirow{2}*{Testing}&Random  & 0.074(0.005)& 0.061(0.010) & 0.033(0.005)&0.011(0.003)\\
&RENES & 0.139(0.019)& 0.120(0.002) & 0.077(0.013)& 0.041(0.002)\\
\bottomrule
\end{tabular}  
\end{table*}

To summarize, for both simple and general cases, we observe that RENES can improve the approximations of NE for different existing solvers, i.e., $\alpha$-rank, CE, FP and PRD. We believe that RENES can be an orthogonal tool for approximating NE in multi-player general-sum games. With combining the advanced solvers and RENES, we can further achieve a better approximation of NE.

\subsection{Ablations}

\begin{figure}[ht]
% \vspace{-10pt}
\centering
\begin{subfigure}[b]{0.225\textwidth}
\centering
\includegraphics[width=0.975\textwidth]{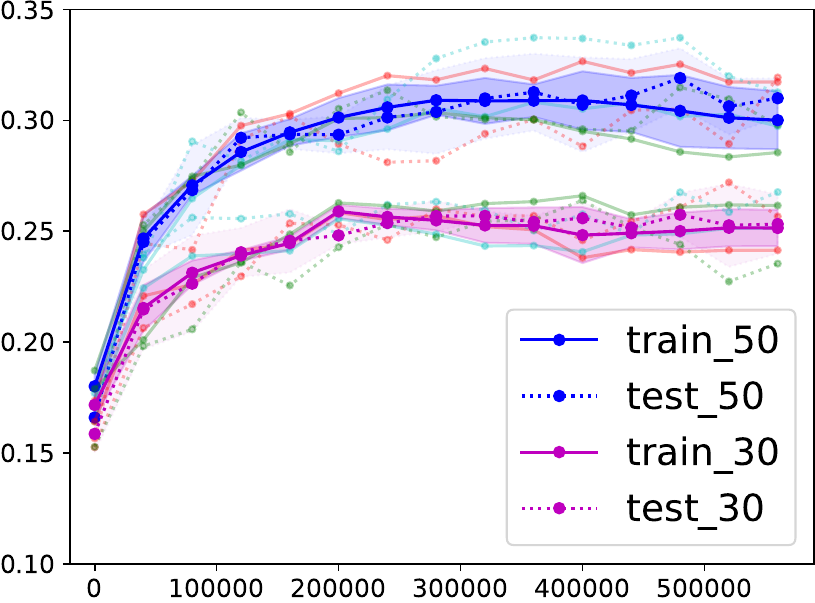}
\caption{Horizon $T$}
\label{fig:simple_abl_step}
\end{subfigure}
\begin{subfigure}[b]{0.225\textwidth}
\centering
\includegraphics[width=0.975\textwidth]{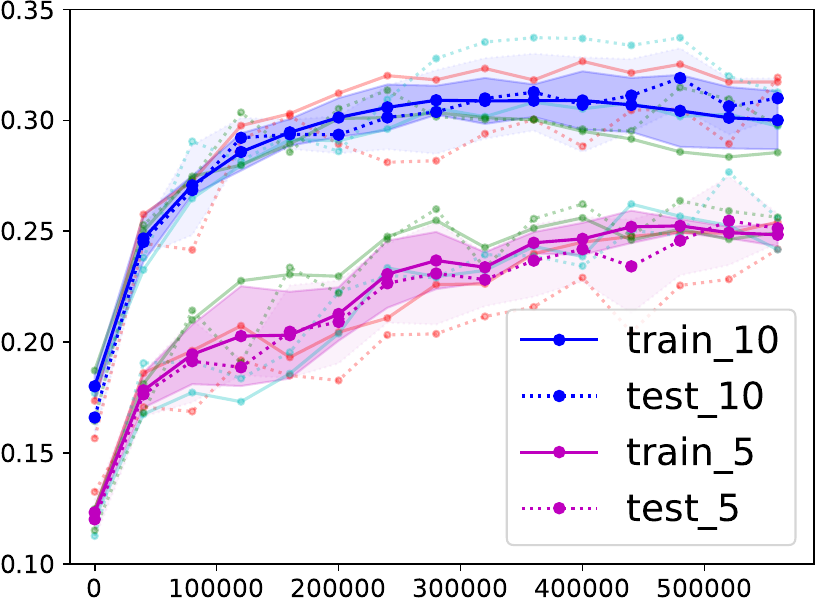}
\label{fig:simple_abl_act}
\caption{Action dimensions $r$}
\end{subfigure}
\caption{Ablations on $\alpha$-rank in simple case.}
\label{fig:simple_ablation}
% \vspace{-10pt}
\end{figure}

We present the ablation results in this section. We ablate two configurations determined by us: i) the number of maximum steps, i.e., horizon $T$, and ii) the number of the action dimension $r$. These two configurations are the main hyperparameters of RENES related to game theory. And for other hyperparameters related to PPO, we do not conduct the ablations. For efficiency, the ablation experiments are conducted on the simple case with $\alpha$-rank as the solver.

The ablation results are displayed in Figure~\ref{fig:simple_ablation}. We observe that with more steps and more actions, the performances are better, which indicates that there is a trade-off between the performance and the efficiency of RENES, and advanced methods can be used for the optimal configurations of the two values given the limited resources, e.g., Optuna~\cite{akiba2019optuna}. We also observe that when the horizon is longer and the number of action dimensions is larger, the results are more sensitive to the seeds, which may due to the intrinsic of the randomness of the initialization of PPO.

\section{Discussion}

\subsection{Limitations and Future Works}

The limitations of RENES are as follows: i) We only conduct experiments on small normal-form games, as for larger games, the running time of RENES, as well as the solvers, will increase. We will consider scaling RENES up to large normal-form games, e.g., 5 players and 30 actions each player, in the future. ii) We only conduct experiments on normal-form games, while extensive-form games (EFGs) are more difficult for computing NE and novel methods to modifying EFGs are required, i.e., the methods to handle the imperfect information and the sequential properties in EFGs. iii) We only focus on NE in this paper. The core idea of modifying the games to facilitate the computation can also be generalized to other solution concepts, e.g., quantal response equilibrium (QRE)~\cite{mckelvey1995quantal}. Due to the space limitation, we provide a detailed discussion about the limitations and future works in Appendix C.

\subsection{Conclusion}

In this work, we propose RENES, which leverages RL methods to find a single policy to modify the original games with different sizes and applies existing solvers to solve the modified games. Our contributions are threefold: i) We adopt the $\alpha$-rank response graph as the representation of the game to make RENES handle the games with different sizes; ii) We leverage the tensor decomposition to improve the efficiency of the modification; iii) We train RENES with the widely-used PPO method. Extensive experiments show that our method can boost the performances of solvers and generate more accurate approximation of NE. To the best of our knowledge, this work is the first attempt to leverage RL methods to train a single policy to modify the games to improve the approximation performances of different solvers in game theory. We hope this method can open a new venue of modifying games as the pre-training task, complimentary to new methods and solutions, to approximate NE with generalizability and efficiency. 

\section*{Acknowledgements}
This research is supported by the National Research Foundation Singapore and DSO National Laboratories under the AI Singapore Programme (AISG Award No: AISG2-GC-2023-009). Hau Chan is supported by the National Institute of General Medical Sciences of the National Institutes of Health [P20GM130461], the Rural Drug Addiction Research Center at the University of Nebraska-Lincoln, and the National Science Foundation under grant IIS:RI \#2302999. Any opinions, findings and conclusions, or recommendations expressed in this material are those of the author(s) and do not reflect the views of the funding agencies.

\section*{Contribution Statement}

Xinrun Wang and Chang Yang are the co-first and co-corresponding authors who contribute to this work equally.

\bibliographystyle{named}
\bibliography{renes}

\clearpage
\appendix

\section{Frequently Asked Questions}
In this section, we want to add more justifications by answering the Frequently Asked Questions (FAQs) from the readers.  

\paragraph{Q1: Why not Use Supervised Learning (SL) Methods?}
The question is briefly discussed in the 4th paragraph of Introduction. We would like to add more discussions here. SL methods are the most straightforward methods to tackle this problem. If using SL methods, we need to generate the training dataset, i.e., randomly sampling games and solving them to obtain NE as labels. Generating datasets is difficult: i) We have to leverage existing methods, e.g., Lemke-Howson methos~\cite{lemke1964equilibrium}, to compute NE, which is time-consuming; ii) We suffer the equilibrium selection problems, where we will have multiple NE, i.e., multiple labels, for one game. Furthermore, the output of SL methods may inevitably be inaccurate and we still need methods to refine the solutions, while SL can not self-improve the solutions without any supervision. 

\paragraph{Q2: The Relation of RENES and Homotopy Methods, e.g., ADIDAS?} Homotopy method is motivated by the topology analysis~\cite{herings2010homotopy}. The basic idea of homotopy is building the continuum between the simple game and the target game and then iteratively approximate the equilibrium of the target game. One recent development of homotopy method is ADIDAS~\cite{gemp2022sample}, which focus on the scalability of the method and the learned policy cannot generalize across different games. However, RENES focus on the generalizatbility, rather than the scalability, and the learned model can be applied to different, even unseen games. Besides, RENES is based on the widely-used PPO method and we will open-source our implementation upon the acceptance of the paper. We note that there is no conflict between RENES and homotopy methods. We expect that homotopy methods can provide the theoretical insights to RENES and guide us to refine the modules in RENES for better performances. 

\paragraph{Q3: The Selection of Baselines.} We want to note that the main objective of RENES is the generalizability, not the scalability or the state-of-the-art (SoTA) performance. Most of the previous methods, e.g., ADIDAS, focus on scalability, i.e., scaling the algorithms to large-scale games. We need to run the algorithms on each specific game and the computed policies do not have the generalizability across different games. While after training, RENES can be applied to any games without training. Therefore, comparing RENES with SoTA methods is unfair, as current SoTA methods require training on specific games. So we only compare with the four widely used solvers. We also want to note that \textbf{RENES can be seamlessly applied to the case where the solver of games is any SoTA method}.

\paragraph{Q4: Plans to Address the Scalability Issues.} One of the main issues of RENES is the scalability, where our experiments focus on demonstrating the generalizability of RENES on small games. We do have plans to address the scalability issues to make RENES scalable to large games, while generalizable. Specifically, our plan is two-fold: i) game abstraction and decomposition, i.e., we will leverage the game decomposition and abstraction techniques to decompose the complex games into a set of small games, where the abstraction is successfully applied to Texas Hold'em poker~\cite{brown2017safe}, and ii) local modification, where instead of global modification, i.e., RENES changes the game globally through changing the weights of the decomposed matrices, we can change the game locally, i.e., given the playing sequence of the game, our model can modify the payoffs of players. 

\paragraph{Q5: Potential Impacts of RENES.} i) The pre-training/fine-tuning paradigm achieves remarkable success in many fields, including natural language processing, computer vision and even reinforcement learning. RENES is in line with this paradigm, i.e., training one policy to modify all games without fine-tuning, and achieves the generalizability across games. ii) Furthermore, our question is that can we build the foundation model for equilibrium finding? The key success of GPT is the pre-text tasks, i.e., next-token prediction. We believe that modifying the games can be a pre-text tasks to train the foundation model for equilibrium finding. \textbf{We believe that for the research in game theory, we should shift from computing equilibria for games by running algorithms on each games to making the models generalizable across different games for efficiency and generalizability.}

\paragraph{Q6: Code Release.} We will release all codes upon the acceptance of the paper. We hope that the code can facilitate the research of the community. 
% \clearpage

\section{More Related Work}
% \section{Related Work}
\label{sec:related_works}
In this section, we provide a concise overview of the four lines of research related to this work, i.e., alternative solutions, learning methods, homotopy methods and combinatorial optimization and discuss the novelties of RENES compared with these methods.

\paragraph{Alternative Solutions.} There are various alternative solutions to facilitate the analysis in game theory. Correlated Equilibrium (CE)~\cite{aumann1974subjectivity,aumann1987correlated} is a solution concept where a mediator can recommend behaviors to players and no player will deviate from the recommended behaviors at the equilibrium. CE can be computed in polynomial time, which is a computationally benign generalization of the intractable NE. Another celebrating result is that there are no-regret learning methods, e.g., regret matching~\cite{hart2000simple}, which can converge to CE. Recently, $\alpha$-rank is proposed as another alternative solution concept to NE~\cite{omidshafiei2019alpha}, which leverages notions from stochastic evolution dynamics in finite populations in the limit of rare mutations. 
% At a high level, $\alpha$-rank models the probability of a population transition from a strategy to a new strategy, by considering the payoff differences between the two strategies. 
The evolution relations between all pairs of strategies are summarized in a response graph, i.e., $\alpha$-rank response graph, and $\alpha$-rank uses the stationary distribution on this graph to characterize the long-term propensity of playing each of the strategies. $\alpha$-rank is unique, which avoids the equilibrium selection problem, and can be computed in polynomial time in general-sum multi-player games~\cite{omidshafiei2019alpha}. 
Some recent works consider the efficient computation of $\alpha$-rank~\cite{yang2020alphaalpha,rashid2021estimating} and use it to characterize the intrinsic properties of games~\cite{omidshafiei2020navigating}. 
Other related solution concepts such as Stackelberg equilibrium~\cite{von2010leadership}, quantal response equilibrium~\cite{mckelvey1995quantal} are not considered in this paper but RENES can be seamlessly applied to these solution concepts with minor modifications.

\paragraph{Learning Methods.} Many learning methods are proposed to approximate NE. Among them, fictitious play (FP)~\cite{brown1951iterative} is one of the most widely-used methods, which has  successfully extended to extensive-form games~\cite{heinrich2015fictitious} and even mean-field games~\cite{perrin2020fictitious}. FP can be viewed as a special case of double oracle~\cite{mcmahan2003planning} or policy space response oracle~\cite{lanctot2017unified}, which utilizes a meta solver to determine the distribution over best-responses, rather than the uniform distribution. Replicator dynamics (RD) is a learning dynamics from evolutionary game theory, which describes a population's evolution via biologically-inspired operators, such as selection and mutation~\cite{taylor1978evolutionary,hennes2020neural}. Projected Replicator Dynamics (PRD) is a specific variant of RD proposed in~\cite{lanctot2017unified}, which enforces the exploration during the learning with RD and shows a superior performance on different games than RD. Many other learning methods such as mirror descent~\cite{sokota2023unified} and counter factual minimization (CFR)~\cite{zinkevich2007regret} can also be the solver in RENES.

\paragraph{Homotopy Methods.} RENES is also related to the homotopy methods~\cite{herings2010homotopy}. In order to solve the game, homotopy methods start constructing a ``simplified'' game with an obvious NE through modifying the payoff values. Then, a continuum between the original game and the simplified game is constructed, where there is a continuous path between the NE in the simplified game and the NE in the original game. Most homotopy methods rely on the exact combinatorial algorithms, e.g., Lemke-Howson method~\cite{herings2001differentiable,herings2002computation,geanakoplos2003nash}, while RENES works on inexact solvers. One recent development of homotopy method is ADIDAS~\cite{gemp2022sample}, which focus on the scalability of the method and the learned policy cannot generalize across different games. However, RENES focus on the generalizatbility, rather than the scalability, and the learned model can be applied to different, even unseen games. Besides, RENES is based on the widely-used PPO method and we will open-source our implementation upon the acceptance of the paper. We note that there is no conflict between RENES and homotopy methods. We expect that homotopy methods can provide the theoretical insights to RENES and guide us to refine the modules in RENES for better performances. 

\paragraph{Combinatorial Optimization.} The computing of NE is closed to the combinatorial optimization. Many recent works tackle combinatorial problems with deep learning~\cite{bengio2021machine,mazyavkina2021reinforcement}. The work~\cite{wang2021bi} is mostly related to our work, where the RL algorithms, e.g., PPO, are used to modify the graphs and heuristic methods are used to solve the modified graphs. We generally follow the schedule and make two main novelties: i) representing the games as $\alpha$-rank response graphs and using graph neural network (GNN) to learn a good representation of the games with different sizes, and ii) decomposing the games with tensor decomposition, which can make the action space fixed for the games with different sizes, to make our methods able to learn a single strategy for games with different sizes.

\section{Limitations and Future Works}

In this section, we provide a more detailed discussion about the limitations and future works. The limitations of RENES are as follows:
\begin{itemize}
    \item The design of the experiments is to demonstrate the generalizability of RENES. We only conduct experiments on small normal-form games, as for larger games, the running time of RENES, as well as the solvers, will increase. We will consider scaling RENES up to large normal-form games, e.g., 5 players and 30 actions each player, in the future. Specifically, our plan to address the scalability is as follows: instead of modifying the game globally, we will let RENES to change the payoff of players locally, i.e., given the play sequences of the games, RENES will modify the players' payoffs. The local modification is similar to the reward shaping~\cite{du2019liir} and our objective is to train a general and foundation model which can provide efficient reward shaping schemes across different games, even unseen games. 
    \item We only conduct experiments on normal-form games, while extensive-form games (EFGs) are more difficult for computing NE and novel methods to modifying EFGs are required, i.e., the methods to handle the imperfect information and the sequential properties in EFGs. We will extend RENES to handle EFGs, e.g., Poker~\cite{brown2018superhuman}, in future works. Specifically, our plan are as follows: the EFGs are usually represented as trees, we will use GNN to represent EFGs to and then generate the modifications of payoffs of players. The main challenge is that normally the EFGs are extremely huge and we cannot full transverse the game, so the efficient sampling method is required.
\end{itemize}

The core idea of modifying the games to facilitate the computation can also be generalized to other solution concepts, e.g., quantal response equilibrium (QRE)~\cite{mckelvey1995quantal}. Continuous games, e.g., generative adversarial networks~\cite{goodfellow2020generative}, are also important and have more relevance in machine learning. We will consider to extend RENES to handle continuous games to further improve the applicability of RENES.

Furthermore, we can take the successful pre-training/fine-tuning framework from computer vision~\cite{bao2021beit,he2022masked} and natural language processing~\cite{kenton2019bert} to game theory, where we can pre-train a modifying strategy of game which can be efficiently fine-tuned to adapt to any solvers, and even any solution concepts, efficiently. As this pre-trained strategy is solver-agnostic and solution-agnostic, we believe it can reveal the intrinsic properties in the game space, as what observed in~\cite{czarnecki2020real,omidshafiei2020navigating,bertrand2022limitations}.

\begin{table}[t]
\caption{Hyperparameters of RENES.} 
\label{tab:hyper}
\centering
\begin{subtable}[t]{.45\textwidth}
\caption{Hyperparameters related to games in RENES}
    \label{tab:hyper_game}
    \centering
\begin{tabular}{c|c}
\toprule
Names     & Values \\
\midrule
Number of games for training     & 3000\\
Number of games for testing & 500 \\
Dimensions of the action $r$ & 10 \\
Horizon $T$ & 50 \\
Payoff range & $[-5, 5]$ \\
Weight step $\eta$ & 5.0 \\
$m$ in $\alpha$-rank response graph & 5.0 \\
$\alpha$ in $\alpha$-rank response graph & 1.0 \\
Node embedding in GNN & 20 \\
Layers of GNN & 2 \\
Hidden size in MLP & 64 \\
Layers of MLP & 3 \\
\bottomrule
\end{tabular}
\end{subtable}
% \hfill
\begin{subtable}[t]{.45\textwidth}
  \centering
\caption{Hyperparameters related to PPO in RENES}
    \label{tab:hyper_ppo}
    \begin{tabular}{c|c}
    \toprule
    Names     & Values \\
    \midrule
     Learning rate    & $1e-3$\\
     Discounting factor $\gamma$ & $0.99$ \\
     Entropy coefficient & 0.01 \\
     Value loss coefficient & 0.5 \\
     Maximum gradient norm & 0.5\\
     Number of processes & 20 \\
     Number of forward steps & 100 \\
     Number of epoch in PPO & 16 \\
     Clip of parameters in PPO & 0.2\\
     Number of mini batch & 64 \\
     Number of environment steps & $6e5$ \\
    \bottomrule
    \end{tabular}
\end{subtable} 
\end{table}
\section{Implementation Details}

\subsection{Games and Solvers}
\label{sec:game_and_solver}
\paragraph{Randomly Sampled Games.} For the games, we randomly sample 3000 games for training and 500 games for testing to verify the ability of RENES to generalize to unseen games. The games are represented as high dimensional payoff tensors with dimensions as $[K, |\mathcal{A}^{1}|, \dots, |\mathcal{A}^{K}|]$, which are used for tensor decomposition. For example, for a 2-player game and each player is with 5 actions, the payoff tensor is a $2\times 5\times 5$ tensor. Without loss of generalizability, the games are scaled into $[-5, 5]$ to avoid any numerical issues when computing the $\alpha$-rank response graph, as well as the inputs to neural networks. We also scale the payoff tensors of the modified games into $[-5, 5]$ at each step during training to stabilize the training process. 

\paragraph{Selected Solvers.} The four selected solvers are: $\alpha$-rank, CE, FP and PRD. For $\alpha$-rank and PRD, we use the implementations in OpenSpiel~\cite{lanctot2019openspiel}\footnote{\url{https://github.com/deepmind/open_spiel}}\footnote{For some randomly sample game, we found that the implementation of $\alpha$-rank cannot work correctly due to the numerical issues. This issue is rarely encountered in the experiments with some random seed, so we try a new seed for the experiment. We do not fix this issues and leave this for future works.}. For the computation, we do not use the solvers provided in~\cite{marris2021multi}, where many criteria, e.g., maximum welfare and maximum entropy, are proposed to do the equilibrium selections,  as they rely on the linear program solvers, e.g., ECOS, provided by \texttt{cvxpy}, which cannot handle \emph{poorly scaled problems}, i.e., values of the objective, constraints, or intermediate results that differ by several orders of magnitude, and report the errors when solving some randomly sampled games. Therefore, we use the regret matching~\cite{hart2000simple} to compute CE, which is also implemented in OpenSpiel. We note that using this method does not explicitly address the equilibrium selection problem and more advanced methods, either learning-based or optimization-based, will be considered in future works. We implement FP by ourselves and verify the correctness on two-player zero-sum games, e.g., RPS game. 
\subsection{Implementation of PPO}
\label{app:ppo_hyperpara}
The implementation of RENES is based on the PPO implementation from~\url{https://github.com/ikostrikov/pytorch-a2c-ppo-acktr-gail}. The values of the hyperparameters are displayed in Table~\ref{tab:hyper}.

\end{document}